\documentclass[fleqn,11pt]{wlscirep}
\usepackage{graphicx}

\newcommand*\rfrac[2]{{}^{#1}\!/_{#2}}
\title{Giant magnetoresistance, three-dimensional Fermi surface and origin of resistivity plateau in YSb semimetal}
\author[1]{Orest Pavlosiuk}
\author[1]{Przemys{\l}aw Swatek} 
\author[1,*]{Piotr Wi\'{s}niewski}
\affil[1]{Institute of Low Temperatures and Structure Research,
Polish Academy of Sciences, PNr 1410, 50-950 Wroc{\l}aw, Poland}
\affil[*]{Corresponding author: p.wisniewski@int.pan.wroc.pl}
\begin{abstract}
Very strong magnetoresistance and a resistivity plateau impeding low temperature divergence due to insulating bulk are hallmarks of topological insulators and are also present in topological semimetals where the plateau is induced by magnetic field, when time-reversal symmetry (protecting surface states in topological insulators) is broken. Similar features were observed in a simple rock-salt-structure LaSb, leading to a suggestion of the possible non-trivial topology of 2D states in this compound. 
We show that its sister compound YSb is also characterized by giant magnetoresistance exceeding one thousand percent and low-temperature plateau of resistivity. We thus performed in-depth analysis of YSb Fermi surface by band calculations, magnetoresistance, and Shubnikov--de Haas effect measurements, which reveals only three-dimensional Fermi sheets. Kohler scaling applied to magnetoresistance data accounts very well for its low-temperature upturn behavior. The field-angle-dependent magnetoresistance demonstrates a 3D-scaling yielding effective mass anisotropy perfectly agreeing with electronic structure and quantum oscillations analysis,  thus providing further support for 3D-Fermi surface scenario of magnetotransport, without necessity of invoking topologically non-trivial 2D states. We discuss data implying that analogous field-induced properties of LaSb can also be well understood in the framework of 3D multiband model.
\end{abstract}
\begin{document}
\flushbottom
\maketitle
\section*{Introduction}
Yttrium monoantimonide has mainly been used as a non-magnetic reference or as a 'solvent' in monoantimonides of $f$-electron-elements solid solutions with anomalous physical properties such as dense Kondo behavior and complex magnetic ground-states\cite{Vogt1993}. Within that context it has been characterized as a metal by low-temperature specific heat measurements\cite{Gambino1971}. Later Hayashi et al. have shown that the first-order phase transition from the NaCl-type to a CsCl-type crystal structure occurs in YSb at 26 GPa \cite{Hayashi2003}. That discovery induced numerous calculations of electronic structure of the compound, among them those by T\"ut\"unc\"u, Bagci and Srivastava, who directly compared electronic structure of YSb with that of LaSb \cite{Tutuncu2007a}. Results of those calculations were very  similar for both compounds, revealing low densities of states at Fermi level and characteristic anti-crossings leading to band inversion at X-points of the Brillouin zone. 
LaSb has a simple NaCl-type structure without broken inversion symmetry, perfect linear band crossing or perfect electron--hole symmetry, yet it exhibits the exotic magnetotransport properties of complex-structure semimetals like TaAs, NbP (Weyl semimetals)\cite{Huang2015a,Shekhar2015a}, Cd$_3$As$_2$ (Dirac semimetal)\cite{Liang2014a} and WTe$_2$ (resonant compensated semimetal)\cite{Ali2014b,Ali2015}.
Recently Tafti et al. discovered in LaSb field-induced resistivity plateau at low temperatures up to $\approx\!15$\,K, ultrahigh mobility of carriers in the plateau region, quantum oscillations, and magnetoresistance ($M\!R$) of nearly one million percent at 9\,T \cite{Tafti2015a}. Their calculations, including spin-orbit coupling (SOC) effect, suggested that LaSb is a topological insulator with a 10\,meV gap open near the X-point of the Brillouin zone. They also observed specific angular dependence of frequencies of quantum oscillations and ascribed them to two-dimensional Fermi surface (FS) possibly formed of topologically nontrivial states, and thus proposed LaSb as a model system for understanding the consequences of breaking time-reversal symmetry in topological semimetals\cite{Tafti2015a}.

However, such angular dependence has already been observed in LaSb by de Haas-Van Alphen measurements and well explained by the presence of elongated pockets of 3D-Fermi surface\cite{Hasegawa1985,Settai1993}. 

Tafti et al. also invoked the opening of insulating gap (i.e. metal-insulator transition) as a source of the field-induced resistivity plateau in LaSb\cite{Tafti2015a} but it should be noted that in the case of WTe$_2$ the existence of a magnetic-field-driven metal-insulator transition has been excluded by means of Kohler scaling analysis of magnetoresistance\cite{Wang2015g}. 

Motivated by these ambiguities in the interpretation of LaSb properties we decided to carry out a comprehensive characterization of magnetotransport properties of a sister compound YSb. We found that YSb displays physical properties in many aspects very similar to those of LaSb. Our results are in accord with those of other groups that appeared during preparation of our article\cite{Ghimire2016,Yu2016}. The interpretation proposed by Ghimire et al. and Yu et al. follows that presented by Tafti et al. for LaSb\cite{Tafti2015a}, implying the role of field-induced metal-insulator transition in YSb. 

However, our analysis of magnetoresistance and Shubnikov--de Haas (SdH) effect provides strong support for 3D-Fermi surface scenario of magnetotransport, without necessity of invoking topologically non-trivial 2D states or metal-insulator transition in YSb.

\section*{Results}
\subsection*{Magnetoresistance and the origin of its plateau}

Electrical resistivity ($\rho$) was measured on two samples (denoted as \#1 and \#2) and its dependence on temperature in zero field is plotted in Supplementary Figure 1(a). Shape of $\rho(T)$ curves is typical for a metal. When measured in different applied fields $\rho(T)$ exhibits universal plateau at temperatures 2--15\,K, as shown for sample \#1 in Supplementary Figure 1(b). Temperature range of this plateau is very similar to that reported for LaSb\cite{Tafti2015a}. 

\begin{figure}[h]
\includegraphics[width=16cm]{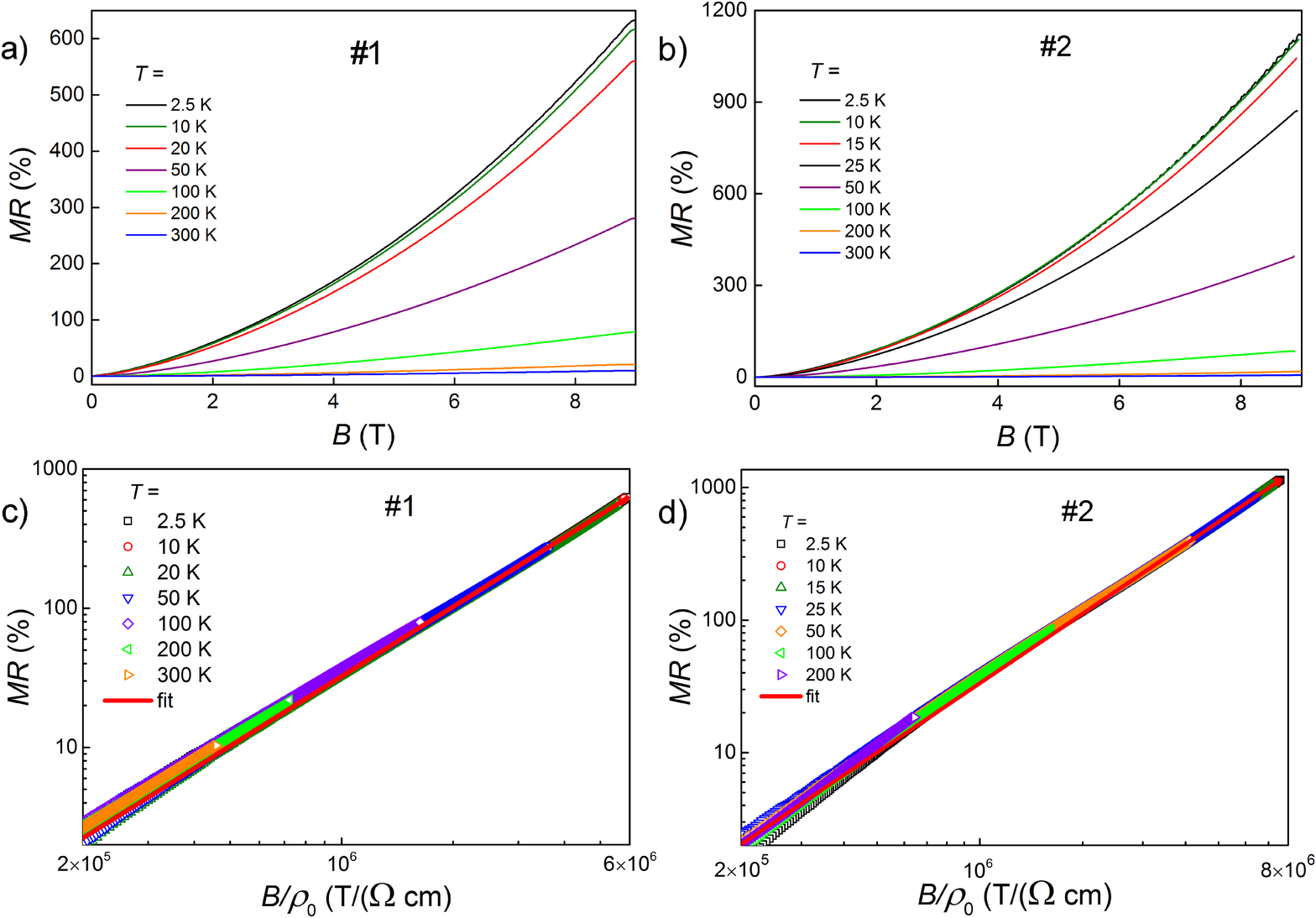}
\caption{(a) and (b)~Magnetoresistance of samples \#1 and \#2, respectively, versus strength of applied magnetic field at different temperatures. (c) and (d)~Kohler scaling of magnetoresistance, $M\!R\!\propto\!(B/\rho_0)^m$ fitted to 2.5\,K data yields $m\!=1.64~{\rm and}~1.74$, for samples \#1 and \#2, respectively.}
\label{MR_Kohler}
\end{figure}
Magnetoresistance, $M\!R\equiv[\rho(B)-\rho(B=0)]/\rho(B=0)$, is plotted versus magnetic field, $B$, in Figs.\,\ref{MR_Kohler}(a) and \ref{MR_Kohler}(b), for samples \#1 and \#2, respectively. Following the approach of Wang et al.\cite{Wang2015g} we performed Kohler scaling analysis of magnetoresistance to test for the existence of a magnetic-field-driven metal-insulator transition in YSb. Figures \ref{MR_Kohler}(c) and \ref{MR_Kohler}(d) show very good Kohler scaling of our data, $M\!R\!\propto\!(B/\rho_0)^m$, yielding at 2.5\,K exponents $m=\,$1.64 and 1.74, respectively, very close for both samples, despite significant difference of their $M\!R$ values. Efficiency of this scaling indicates that resistivity plateau is due purely to the magnetoresistance, but not to a field-induced metal-insulator transition. 

\begin{figure}[h]
\includegraphics[width=16cm]{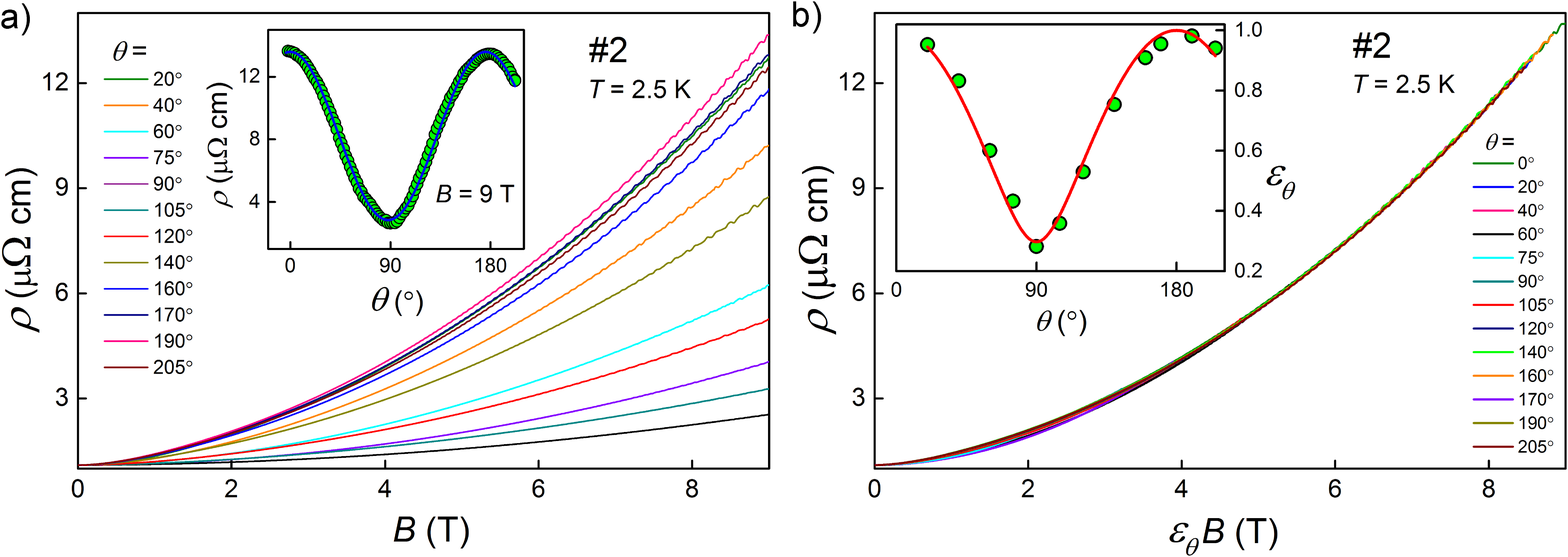}
\caption{(a) Resistivity of YSb (sample \#2) at 2.5\,K versus strength of  magnetic field applied at different angles $\theta$. Inset: resistivity at 2.5\,K and in 9\,T versus field rotation angle; blue line represents $\rho\propto\cos\theta$ dependence. (b) Data of (a) replotted with $B$ scaled by angle-dependent factor $\varepsilon_\theta$. Inset: angle dependence of $\varepsilon_\theta$; red line represents fit with $\varepsilon_\theta=(\cos^2\theta+\gamma^{-2}\sin^2\theta)^{1/2}$ function yielding mass anisotropy $\gamma=3.4$.}
\label{rho_angle}
\end{figure}

In order to elucidate the dimensionality of FS we measured  magnetoresistance of the sample \#2 in fields applied at different angles to its surface, Fig.\,\ref{rho_angle}(a). Here $\theta=0^\circ$ denotes the field perpendicular to the sample surface and the current direction, whereas $\theta=90^\circ$ means that the field is parallel to the current. As shown in the inset to Fig.\,\ref{rho_angle}(a), $\rho$ at strongest field of 9\,T follows $\propto\cos\theta$. This is typical behavior for materials without magnetic anisotropy, but the change of $\rho$ expressed as anisotropic magnetoresistance, $A\!M\!R\!\equiv\![\rho(90^\circ)\!-\!\rho(0^\circ)]/\rho(0^\circ)$, has an outstanding -80\% value.   

Moreover, when field strength is scaled by a factor $\varepsilon_\theta$ dependent on mass anisotropy and $\theta$-angle, all $\rho(T)$ data of Fig.\,\ref{rho_angle}(a) collapse on single curve, as shown in Fig.\,\ref{rho_angle}(b). 
Inset of Fig.\,\ref{rho_angle}(b) shows that values of $\varepsilon_\theta$ plotted against field angle $\theta$ can be perfectly fitted with $\varepsilon_\theta\,=(\cos^2\theta+\gamma^{-2}\sin^2\theta)^{1/2}$ function, shown with red line. Such scaling has initially been proposed for anisotropic superconductors\cite{Blatter1992}, and recently used to interpret $M\!R$ behavior of WTe$_2$ based on its 3D electronic nature\cite{Thoutam2015}. Parameter $\gamma$ represents effective mass anisotropy of carriers mostly contributing to the magnetoresistance.
We ascribe this behavior to a strongly anisotropic sheet of 3D-FS revealed by SdH effect data, as shown below.

\subsection*{Hall effect}
\begin{figure}
\includegraphics[width=17cm]{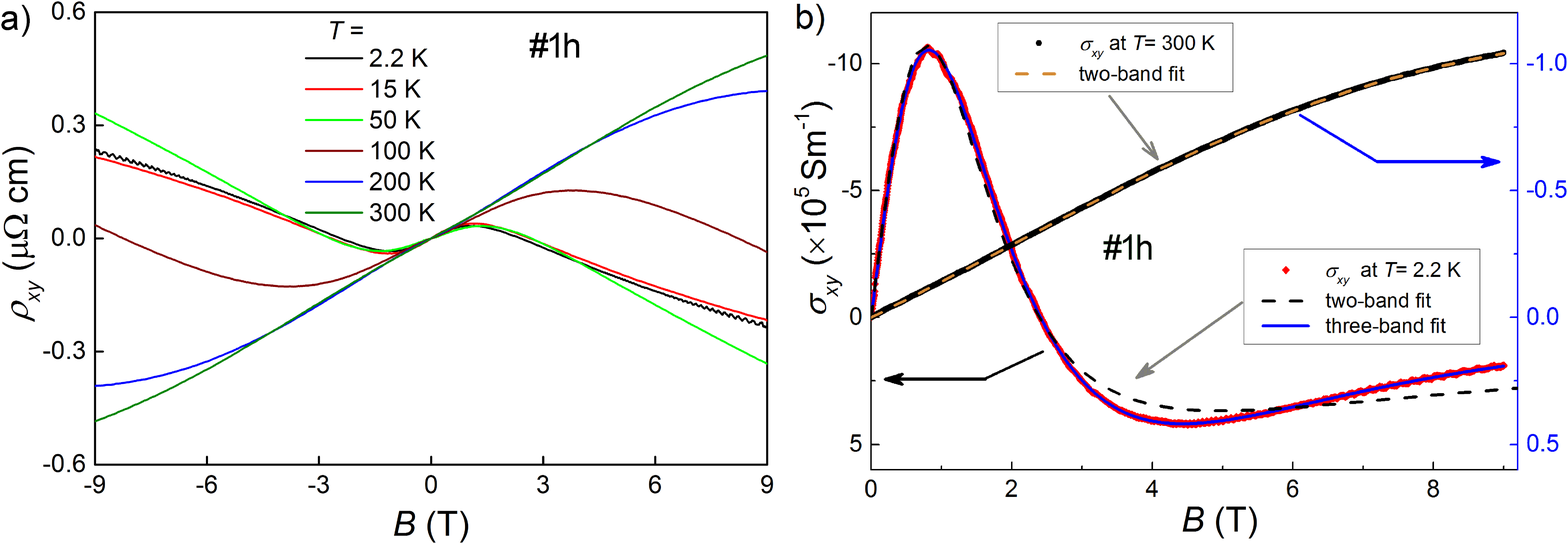}
\caption{(a) Hall resistivity of YSb (sample \#1h) versus magnetic field recorded at several temperatures. (b) Magnetic field dependent Hall conductivity at $T=2.2$\,K (left axis) and 300\,K (right axis). Lines represent the fits with multiple-band model (Eq.\,\ref{two-band-sigma-model}).}
\label{rho_xy}
\end{figure}
\par Hall resistivity of sample \#1h (cut from the same single crystal as \#1) measured at several temperatures between 2.2\,K and 300\,K is shown in Fig.\,\ref{rho_xy}(a). Nonlinear $\rho_{xy}(B)$ indicates that at least two types of charge carriers are responsible for the Hall effect observed in YSb.  
The $\rho_{xy}(B)$ curves for temperatures from 2.2--15\,K range are almost identical, which points to nearly constant carrier concentrations and mobilities, and coincides with the plateau of $\rho_{xx}(T)$ observed in the same range of $T$. Changes of sign of $\rho_{xy}(B)$ observed at $T\leq100$\,K, indicate conducting bands of both electrons and holes, at higher temperatures $\rho_{xy}(B)$ is positive in the whole range of magnetic field ($0<B\leq9$T). Thus, the Hall contributions of holes and electrons nearly compensate, but both depend on temperature in different manner. Clear quantum oscillations are observed in $\rho_{xy}$ in temperature range 2.2--15\,K (cf. Figs.\,\ref{rho_xy}(a) and (b)). 
\par 
Since $\rho_{xy}\ll\rho_{xx}$, the off-diagonal component of conductivity tensor $\sigma_{xy}=-\rho_{xy}/(\rho_{xx}^2+\rho_{xy}^2)$ should be used for multiple-band analysis of Hall data. In this case simple Drude model can be used: 
\begin{equation}
\sigma_{xy}(B)=eB\sum_{i}\frac{n_i \mu_i^2}{1+(\mu_iB)^2}, 
\label{two-band-sigma-model}
\end{equation}
summing up conductivities of individual bands, with $n_i$ and $\mu_i$ denoting respectively concentration and mobility of carriers from the $i$-th band\cite{Hurd1972}. As Fig.\,\ref{rho_xy}(b) shows, for data collected at 300\,K accounting for two bands yielded a good fit. On the other hand, fitting with two bands was insufficient for 2.2\,K data, but addition of a contribution of another band with small concentration of more mobile holes brought a very satisfactory fit. 
\par 
For LaSb Tafti et al.\cite{Tafti2015a} estimated uncompensated carrier concentration, $n$, and the Hall mobility ${\rm\mu_H}$, using the relations $n=1/eR_{\rm H}(0)$ (with $R_{\rm H}(0)$ being the zero-temperature limit of $R_{\rm H}(T)$) and $\mu_H=R_{\rm H}(0)/\rho_0$. They obtained $n\approx10^{20}{\rm\,cm}^{-3}$ and ${\rm\mu_H\approx10^5\,cm^2/(Vs)}$. 
\par
For our YSb sample the fit with  with multiple-band model (Eq.\,\ref{two-band-sigma-model}), shown in Fig.\,\ref{rho_xy}(b), yielded at $T=$2.2\,K the concentrations: $n_e=1.52\!\times\!10^{20}{\rm\,cm}^{-3}$,
$n_h=1.16\!\times\!10^{20}{\rm\,cm}^{-3}$, and mobilities 
$\mu_e=2.7\!\times\!10^3{\rm\, cm^2/(Vs)}$, 
$\mu_h=1.9\!\times\!10^3{\rm\, cm^2/(Vs)}$. Third band necessary for that fit consists of more mobile holes [$n=3.4\!\times\!10^{19}{\rm\,cm}^{-3}$, $\mu=7.9\!\times\!10^3{\rm\, cm^2/(Vs)}$]. These results are consistent with band calculations presented below. 

The fit to data collected at 300\,K yielded: the concentrations: 
$n_e=1.34\!\times\!10^{18}{\rm\,cm}^{-3}$, $n_h=1.94\!\times\!10^{19}{\rm\,cm}^{-3}$, and mobilities $\mu_e=1.8\!\times\!10^3{\rm\,cm^2/(Vs)}$, $\mu_h=8.2\!\times\!10^2{\rm\,cm^2/(Vs)}$, so
YSb has similar concentration of carriers but with significantly lower mobility than LaSb. This seems to be the main reason for its significantly smaller magnetoresistance, as it was well demonstrated for WTe$_2$.\cite{Ali2015}  

Overall, Hall effect results are in perfect agreement with characteristics of Fermi surface presented in the next section.
\begin{figure}[t]
\includegraphics[width=15cm]{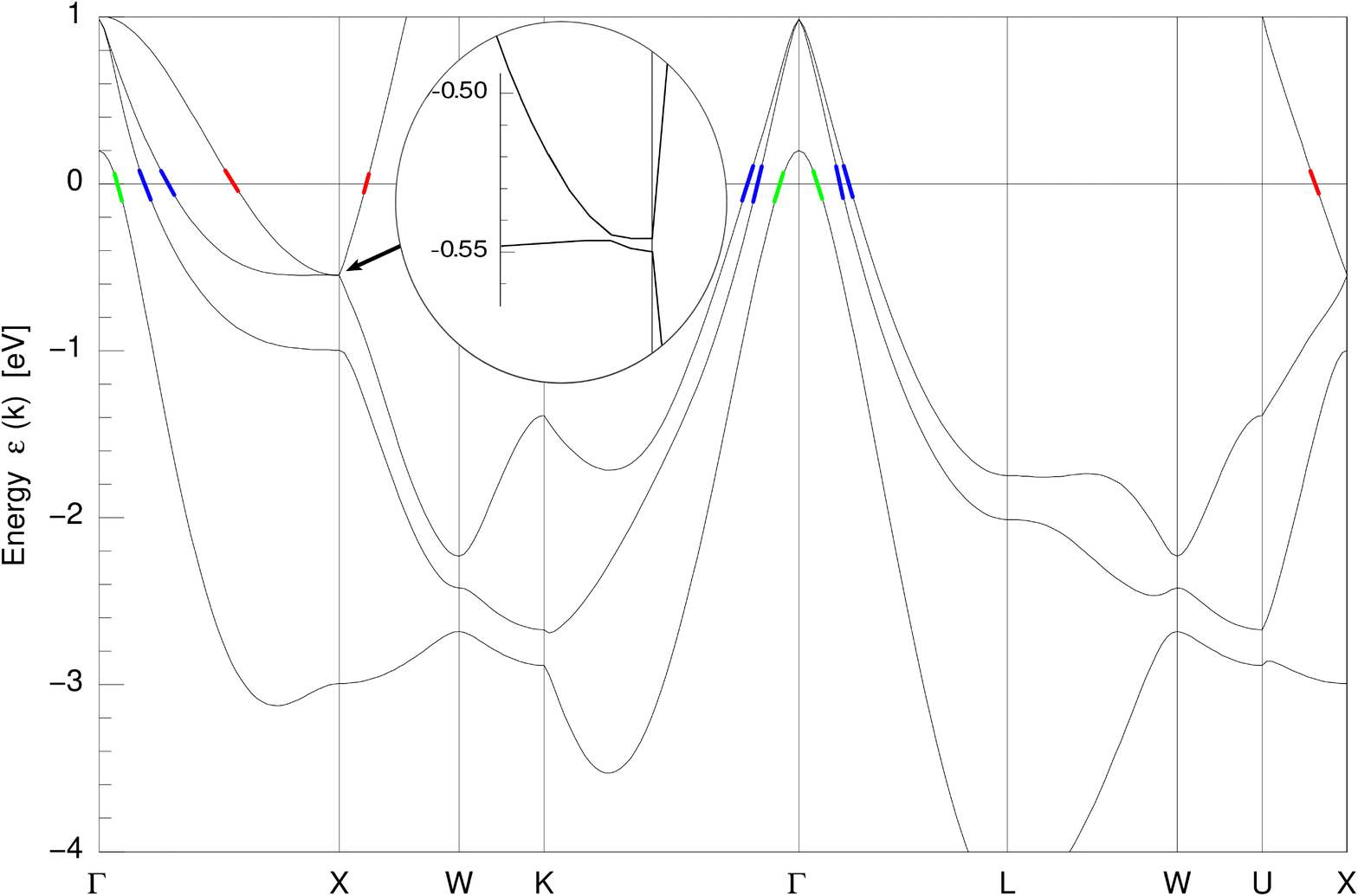}
\caption{Relativistic energy band structure for YSb calculated using FPLO approach with GGA approximation. Bands crossing Fermi level are marked in colors: electron band $\alpha$ in red, hole bands: $\beta$ and $\delta$ in blue, $\zeta$ (not observed in SdH oscillations) in green. Vicinity of X-point with an anti-crossing and opened gap is shown as blow-up in inset.}
\label{BandStr}
\end{figure}
\subsection*{Fermi surface analysis: electronic structure calculations and Shubnikov-de Haas effect}
\begin{figure}
\includegraphics[width=17cm]{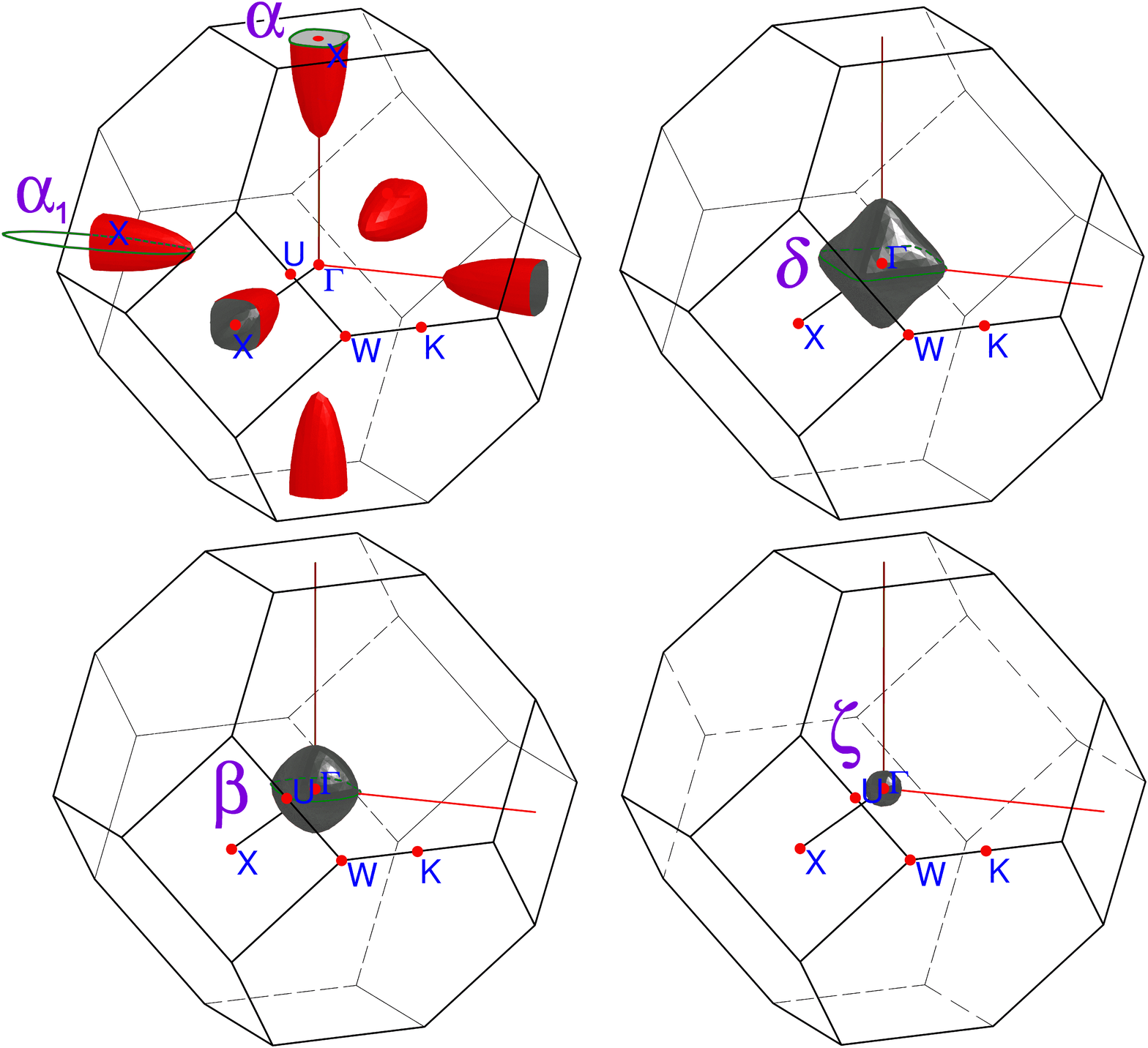}
\caption{Fermi surface of YSb from electronic band calculations. Green lines indicate cyclotron orbits (extreme cross-sections of FS-sheets $\alpha, \beta, \delta$ and $\alpha_1$), for which we observed SdH oscillations in fields applied at $\theta=0^\circ$ (cf.~Fig.\,\ref{FFT-spectrum} and Table\,\ref{LMtable}).}
\label{FSsheets}
\end{figure}

We performed the electronic structure calculations for YSb using a full potential all-electron local orbital code (FPLO) within GGA approximation. Figure\,\ref{BandStr} shows obtained energy band structure for YSb, with a few bands crossing Fermi level. Near X-point there is an anti-crossing present with a gap of $\approx0.8{\rm\,meV}$, similar to those reported for lanthanum monopnictides,  which led to the proposal of 2D topologically non-trivial states at the origin of extraordinary behavior of their magnetoresistance\cite{Zeng2015,Tafti2015a}.
Fermi surface (FS) resulting from our calculations is presented in Fig.\,\ref{FSsheets}: two electron sheets centered at X-points and three nested hole sheets centered at $\Gamma$-point. Our calculations are in good agreement with those of Ref.\,\cite{Ghimire2016}.

\begin{figure}
\includegraphics[width=14cm]{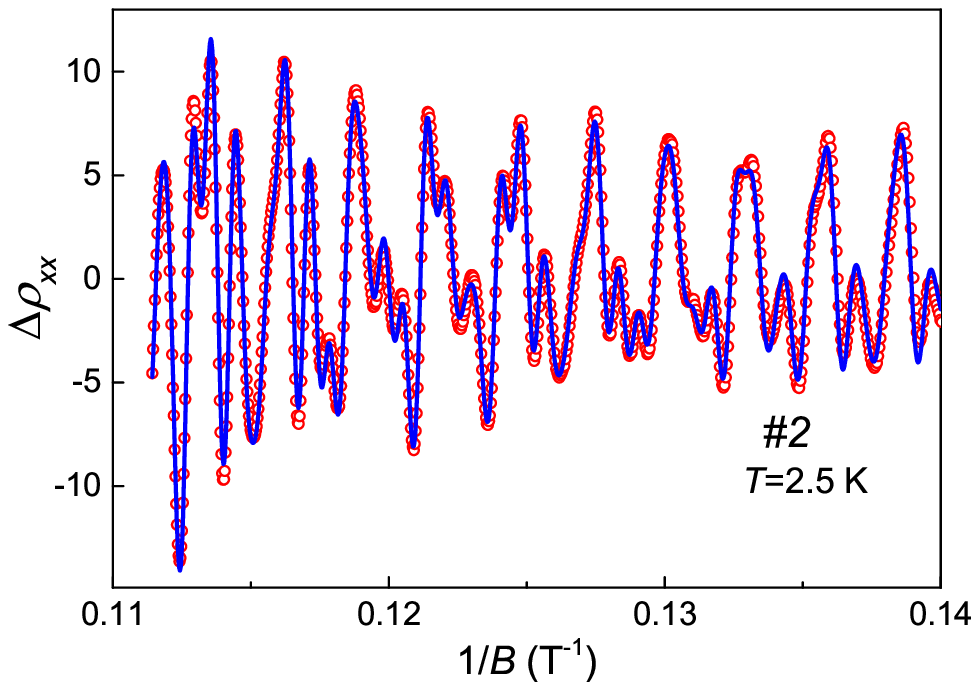}
\caption{Shubnikov-de Haas oscillations of resistivity at $T=2.5$\,K. Blue solid line represents the fit with multiple Lifshitz-Kosevich function (Eq.\,\ref{multiLK}).}
\label{SdH-LK-fit}
\end{figure}

\begin{figure}
\includegraphics[width=17cm]{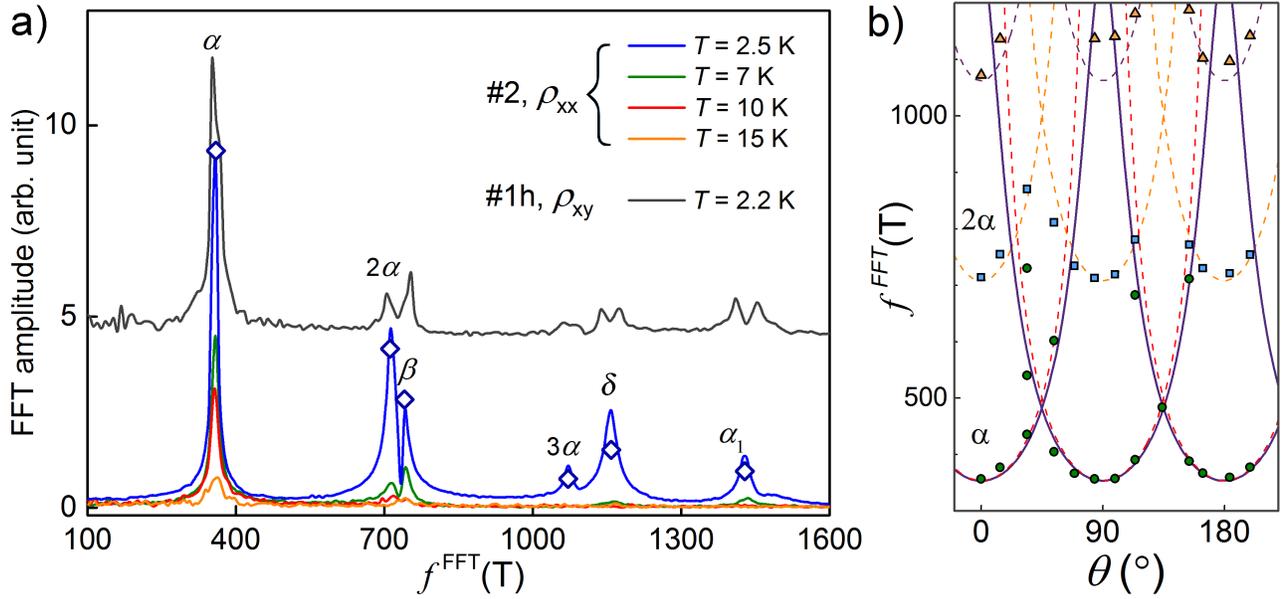}
\caption{(a) FFT-frequency spectrum of $\rho_{xx}(1/B)$ oscillations of sample \#2 measured at different temperatures. Diamond symbols indicate normalized amplitudes obtained from fitting with multiple Lifshitz-Kosevich function shown in Fig.\,\ref{SdH-LK-fit}. Black solid line (for the sake of clarity offset by 5 units) represents the spectrum of $\rho_{xy}(1/B)$ oscillations of sample \#1h at 2.5\,K.
(b) Frequencies of two strongest FFT components of SdH oscillations shown in Fig.\,\ref{rho_angle}(a): $\alpha$ (circles) and 2$\alpha$ (squares), as well as 3$\alpha$ (triangles), versus sample rotation angle $\theta$. Dashed lines represent $f\propto\cos^{-1}\theta$, solid line corresponds to $f=f_0(\sin^2\theta+r^2\cos^2\theta)^{-1/2}$ with $r=3.6$. This line is redrawn for $\theta\!+\!\pi/2$, reflecting the symmetry of cubic YSb lattice.}\label{FFT-spectrum}
\end{figure}

Shubnikov--de~Haas oscillations of resistivity are discernible for YSb at temperatures up to 15\,K and in fields above 6\,T, as seen in Figs.\,\ref{MR_Kohler}(a), \ref{MR_Kohler}(b) and \ref{rho_angle}(a). 
Since $\rho_{xy}\ll\rho_{xx}$, we may safely assume that the conductivity $\sigma_{xx}\simeq\rho_{xx}^{-1}$ and analyze directly oscillations of $\rho_{xx}$. Points in Fig.\,\ref{SdH-LK-fit} represent resistivity of sample \#2 measured at 2.5\,K, after subtraction of smooth background, plotted versus inverse field $1/B$ (for $7<B<9\,$T). Complex shape of $\Delta\rho_{xx}(1/B)$ dependence indicates that observed SdH oscillation has several components. Indeed fast Fourier transform (FFT) analysis reveals clearly six well separated frequencies (Fig.\,\ref{FFT-spectrum}(a)). 
\par
We chose to fit $\Delta\rho_{xx}(1/B)$ with the multi-frequency Lifshitz-Kosevich function\cite{Lifshitz1958,Shoenberg1984,Seiler1989} (Eq.\,\ref{multiLK}) because, as the maxima of total $\Delta\rho_{xx}$ do not correspond to the maxima of particular components with different frequencies, it is inadequate to determine phases in a multicomponent SdH oscillation using the so-called Landau-level fan diagram (plot of the values of $1/B_N$ corresponding to the $N$-th maximum in $\Delta\rho_{xx}$ versus $N$).  
\begin{equation}\Delta\rho_{xx} = \sum_{i} a_i\sqrt{1/B}\:
\frac{\exp(-c_i/B)}{\sinh(b_i/B)}\:
\cos\bigg(2\pi\big(f_i/B-\varphi_i-\frac{1}{8}\big)\bigg),
\label{multiLK}\end{equation} 
where for each $i$-th component: $f_i$ is the frequency, $\exp(-c_i/B)$ represents the Dingle factor $R_D$ and $a_i\sqrt{1/B}/\sinh(b_i/B)\;[\:\propto \sqrt{B}\:R_TR_S]$ comprises the temperature reduction factor $R_T$ and the spin factor $R_S$. Detailed form of Eq.\,\ref{multiLK}, taking into account harmonic components, and description of $R_D$, $R_T$ and $R_S$ is presented in Supplementary Material.  

The fit including six components was of very good quality, as shown by blue line in Fig.\,\ref{SdH-LK-fit}. Obtained parameters are collected in Table~\ref{LMtable}. All frequencies converged almost exactly to $f_i^{FFT}$ values obtained from FFT analysis. We ascribe oscillations with frequencies of 720 and 1072\,T to second and third harmonic of the strongest one with $f_i=360$\,T.
The phases $\varphi_i$ of fundamental oscillations resulting from the fit are close to Onsager phase factor of $\rfrac{1}{2}$ expected for free electrons. Thus all components of SdH oscillation can be assigned to 3D-FS sheets predicted by band calculations (shown above in Fig.\,\ref{FSsheets}) and no Berry phase of $\pi$ was observed, which could reveal topologically non-trivial charge carriers. 

\begin{table}[h]
\caption{Parameters obtained from fit of the multiple Lifshitz-Kosevich function (Eq.\,\ref{multiLK}) to data-points shown in Fig.\,\ref{SdH-LK-fit}. Frequencies derived by FFT analysis of SdH oscillations of $\rho_{xy}$ are shown for comparison.}\label{paramsLK}
\centering
\begin{tabular}{l *{6}{c}} \hline\hline 
{~~~~$i=$}&$\alpha$&$2\alpha$&$\beta$&$3\alpha$&$\delta$&$\alpha_1$ \\\hline
$f_i^{FFT}\!=\!f_i\!\:\;({\rm T})$& 360 & 720 & 740 & 1072 & 1160 & 1430\\
$f_i^{FFT}|\rho_{xy}\:\:\:({\rm T})$& 359 & 708 & 753 & 1066,\,1082~~~& 1138,\,1174~~~& 1411,\,1452\\
$\varphi_i$ & 0.74(3)& 0.477(4) & 0.69(2) & 0.53(1) & 0.69(1) & 0.31(1)\\ 
\hline\hline
\end{tabular}\label{LMtable}
\end{table}

We performed FFT analysis for all data sets presented in Fig.\,\ref{rho_angle}(a), which allowed us to observe angular behavior of frequencies corresponding to all extreme cross-sections of Fermi sheets and compare them to those derived from our band structure calculations, as shown in Fig.\,\ref{FFT-spectrum}(a). Three of observed frequencies were clearly changing upon rotation of the magnetic field: the principal, labeled as $\alpha$ ($f_{\alpha}^{FFT}=360\,$T at $\theta=0^\circ$), and its harmonics, $2\alpha$ and $3\alpha$. These frequencies are plotted versus $\theta$ in Fig.\,\ref{FFT-spectrum}(b). 
\par
It became apparent from the shape of FS obtained from band calculations (cf. Fig.\,\ref{FSsheets}) that angular behavior of $f_{\alpha}^{FFT}$ follows a cross-section area of a prolate ellipsoid (which well approximates the shape of electron sheet centered at X-point shown in Fig.\,\ref{FSsheets}) by the ((1-$\cos\theta)\;0\;1)$ plane passing by X-point. When magnetic field is applied along $[0\;0\;1]$ direction the plane is perpendicular to $\Gamma$-X line. When field is tilted from $[0\;0\;1]$ towards $[1\;0\;0]$ by the angle $\theta$ the cross-section increases as $S(\theta)=\pi k_x^2(\sin^2\theta+r^2\cos^2\theta)^{-1/2}$. This holds for a two-axial ellipsoid described by the equation: $(x/k_x)^2+(y/k_x)^2+(z/k_z)^2=1$, with $r=k_z/k_x$. Band structure shown in Fig.\,\ref{BandStr} yields $r\approx\,3.6$ (with $k_x$ estimated as average size of $\alpha$-sheet of FS along X-U and X-W lines, and $k_z$ as its size along $\Gamma$-X line). 
\par
After rotation by $\theta=90^\circ$ the $\alpha$ frequency meets the one denoted as $\alpha_1$, initially (i.e. at $\theta=0^\circ$) corresponding to the largest cross-section of the same ellipsoidal FS sheet. Two other observed frequencies, $\beta$ and $\delta$ do not change notably with $\theta$, as expected for almost isotropic hole Fermi sheets centered at $\Gamma$-point. 
\par
We plotted $S(\theta)$ (for $r$=3.6) in Fig.\,\ref{FFT-spectrum}(b) as solid lines. The $\propto\!\cos^{-1}\theta$ dependence, expected for two-dimensional FS, behaves similarly and is shown for comparison with dashed lines. The $\propto\!\cos^{-1}\theta$ dependence was used by Tafti et al.\cite{Tafti2015a} as a hint of possible topologically-nontrivial states in LaSb. However, Fermi surface of that compound has already been well characterized by band calculations and angle-dependent de Haas--van Alphen measurements\cite{Hasegawa1985,Settai1993}, revealing FS very similar to the one we found in YSb, namely consisting of one ellipsoidal electron sheet centered at X-point and two isotropic hole pockets centered at $\Gamma$-point. Hasegawa\cite{Hasegawa1985} and Settai et al.\cite{Settai1993} assigned angular dependence of the principal de Haas--van Alphen frequency (identical to the SdH frequency in Ref.~\cite{Tafti2015a}) to the cross-section of the ellipsoidal sheet $S(\theta)$ in accord with our interpretation of $S(\theta)$ behavior for YSb. This underscores the similarities between these two compounds and implies that there is no need to invoke topologically non-trivial states to explain exotic magnetotransport properties neither in YSb nor in LaSb (contrary to Ref.~\cite{Tafti2015a}). 
\section*{Discussion and conclusions} 
YSb is another material displaying giant magnetoresistance (1100\,\% in 9\,T), three orders of magnitude smaller than that of sister compound LaSb\cite{Tafti2015a}, thus it cannot be termed 'extreme magnetoresistance' (XMR).
This is due mainly to its lower carrier mobility and weaker electron-hole compensation revealed by our Hall effect measurement. 

Kohler scaling analogous to that shown in Figs.\,\ref{MR_Kohler}(c) and (d) has recently been used to explain the remarkable up-turn behavior of $M\!R$ in WTe$_2$ without the field-induced metal-insulator transition or significant contribution of an electronic structure change\cite{Wang2015g}. The same authors have shown that perfect carrier compensation leads to exponent $m=\,$2 in this scaling. 
Kohler scaling for YSb yielded  for our samples the exponents $m=\,$1.64 and 1.74, which seems related to weaker carrier compensation than nearly perfect one in WTe$_2$, where $m=\,$1.92 \cite{Wang2015g}. Thus, analogously to WTe$_2$, Kohler scaling indicates that the field-induced metal-insulator transition is unnecessary to explain up-turn and low-temperature plateau of resistivity in YSb. The origin of the up-turn is a combination of magnetoresistance with the low-temperature resistivity plateau present already at zero field.  
Given the similarity of YSb and LaSb the same may also be true for the latter compound.

Comparing results of SdH measurements with those of electronic structure calculations we obtained comprehensive description of the Fermi surface of YSb. Presence of both electron and hole sheets of similar volumes provides partial charge compensation responsible for its strong magnetoresistance. Band structures of YSb and LaSb are very similar. All Fermi sheets in YSb but the smallest one centered at $\Gamma$-point have their counterparts in LaSb.\cite{Hasegawa1985,Settai1993,Tafti2015a} Our analysis of angular behavior of SdH frequencies in YSb indicates it is related to the three-dimensional FS, in line with Hasegawa\cite{Hasegawa1985} and Settai et al.\cite{Settai1993} findings for LaSb, but not connected to possible non-trivial topology of electronic structure analogous to that suggested by Tafti et al.\cite{Tafti2015a} 

Angular behavior of $M\!R$ can also be perfectly explained by anisotropy of 3D-FS. When field strength is scaled by the angle-dependent factor $\varepsilon_\theta$, all data of Fig.\,\ref{rho_angle}(a) collapse on single curve. 
The effective mass anisotropy factor $\gamma\,=3.4$, obtained from the fit of $\varepsilon_\theta(\theta)$ with the expression $(\cos^2\theta+\gamma^{-2}\sin^2\theta)^{1/2}$, is in excellent agreement with $k_z/k_x\,=3.6$ we estimated  for $\alpha$-sheet of FS. 
This is not surprising, since the mass anisotropy directly reflects the shape of FS, but it shows that angular behavior of $M\!R$ in YSb is mainly governed by anisotropic form of $\alpha$-sheet of FS. That sheet corresponds to the electron band, all other FS-sheets contain holes and are nearly isotropic. The effective mass and mobility of $\alpha$-sheet electrons change significantly with field angle, which strongly modifies the magnetoresistance. 
\par
It has been proposed that the magnetic field induces the reconstruction of the FS in a Dirac semimetal by breaking the time reversal invariance\cite{Young2012,Wang2012b,Burkov2011,Liu2014e}. Assisted by the high mobility of carriers such reconstruction has been suggested to induce very large $M\!R$ observed in Cd$_3$As$_2$ and NbSb$_2$ \cite{Wang2013c,Wang2014}. We also observe features in the electronic structure of YSb, buried under the Fermi level, which may possibly allow the magnetic field to transform this compound into Dirac semimetal. A small gap between inverted bands near the X-point (cf. inset to Fig.\,\ref{BandStr}) might result in topologically non-trivial states.  The effect of FS reconstruction could be similar to temperature-induced Lifshitz transition in WTe$_2$ \cite{Wu2015}, whereas its mechanism might be related, for example, to that of Lifshitz transition driven by magnetic field in CeIrIn$_5$ \cite{Aoki2016a}. 
Very recently Dirac states have been observed by angle-resolved-photoemission spectroscopy in NbSb, a compound with bulk electronic structure very similar to that of YSb\cite{Neupane2016a}, however topologically protected states were not detected in YSb by this method.\cite{He2016a} 
 
Although a small contribution of topologically non-trivial 2D states cannot be completely excluded our analysis of magnetoresistance and Shubnikov--de Haas effect provides strong support for 3D-Fermi surface scenario of magnetotransport properties in YSb. Analogous field-induced properties of LaSb can most probably  be also described in the framework of 3D multiband model. 
\section*{Methods}
Measurements were performed using a Physical Property Measurement System (Quantum Design) on two samples cut from one single crystal and labeled as \#1 and \#1h, and a sample cut from another single crystal and labeled \#2. All samples had shapes of rectangular cuboid with all edges along $\langle1\;0\;0\rangle$ crystallographic directions. Their sizes were: $0.56\!\times\!0.25\!\times\!0.12\;{\rm mm}^3$, $0.4\!\times\!0.47\!\times\!0.13\;{\rm mm}^3$ and $0.41\!\times\!0.32\!\times\!0.09\;{\rm mm}^3$, for samples \#1, \#1h and \#2, respectively. The electric current was always flowing along $[1\;0\;0]$ crystallographic direction. Single crystals were grown from Sb flux and their NaCl-type crystal structure was confirmed by powder X-ray diffraction carried out using an X’pert Pro (PANanalytical) diffractometer with Cu-Ka radiation. No other phases were detected and lattice parameter of 6.163\,\AA\, was determined, reasonably close to literature value 6.155\,\AA \cite{Hayashi2003}. 
Electronic structure calculations were carried out using FPLO-9.00-34 code within generalized gradient approximation (GGA) method\cite{Koepernik1999}. The full-relativistic Dirac equation was solved self-consistently, treating exactly all relativistic effects, including the spin-orbit interaction without any approximations. The Perdew--Burke--Ernzerhof exchange-correlation potential\cite{Perdew1996} was applied and the energies were converged on a dense $k$ mesh with 24$^3$ points. The convergence was set to both the density (10$^{-6}$ in code specific units) and the total energy (10$^{-8}$ Hartree). For the Fermi surface a 64$^{3}$ mesh was used to ensure accurate determination of the Fermi level.\newpage

\section*{Acknowledgment}
We are very grateful to Dr. Zhili Xiao for discussion. This work was supported by the National Science Centre of Poland, grant no. 2015/18/A/ST3/00057. The band structure calculations were carried out at the Wroc{\l}aw Centre for Networking and Supercomputing, grant no. 359.
\section*{Author contributions} 
O.P. conducted all the experiments, O.P. and P.W. conceived the experiments and analyzed their results, P.S. carried out the electronic structure calculations and prepared Figures 5 and 6. All authors reviewed the manuscript. 
\newpage
\newcommand{\fSref}[1]{Fig.~S\ref{#1}}
\setcounter{figure}{1}
\rfoot{Supp. Mat. page 1 of 3}
\noindent{\Large{\bf SUPPLEMENTARY MATERIAL}}\\\\
{\Large{\bf Giant magnetoresistance, three-dimensional Fermi surface\\ and origin of resistivity plateau in YSb semimetal}}\\\\
{\large Orest Pavlosiuk, Przemys{\l}aw Swatek and Piotr Wi\'{s}niewski}\\\\
{\small Institute of Low Temperatures and Structure Research,
Polish Academy of Sciences, Wroc{\l}aw, Poland}
\section*{Resistivity and its universal low-temperature plateau} 
In Fig.\,S1(a) we show metallic-like temperature dependence of electrical resistivity $\rho$ of samples \#1 and \#2. The room- to residual-resistivity ratio, $R\!R\!R\;(\equiv\rho(300\,{\rm K})/\rho(2\,{\rm K}))$, is 13 and 22, respectively. Behavior of $\rho(T)$ of sample \#1 in different applied magnetic fields is shown in Fig.\,1(b). Plateau at $T\leq15\,$K is observed independent of strength of applied field.
\begin{figure}[h]
\includegraphics[width=10cm]{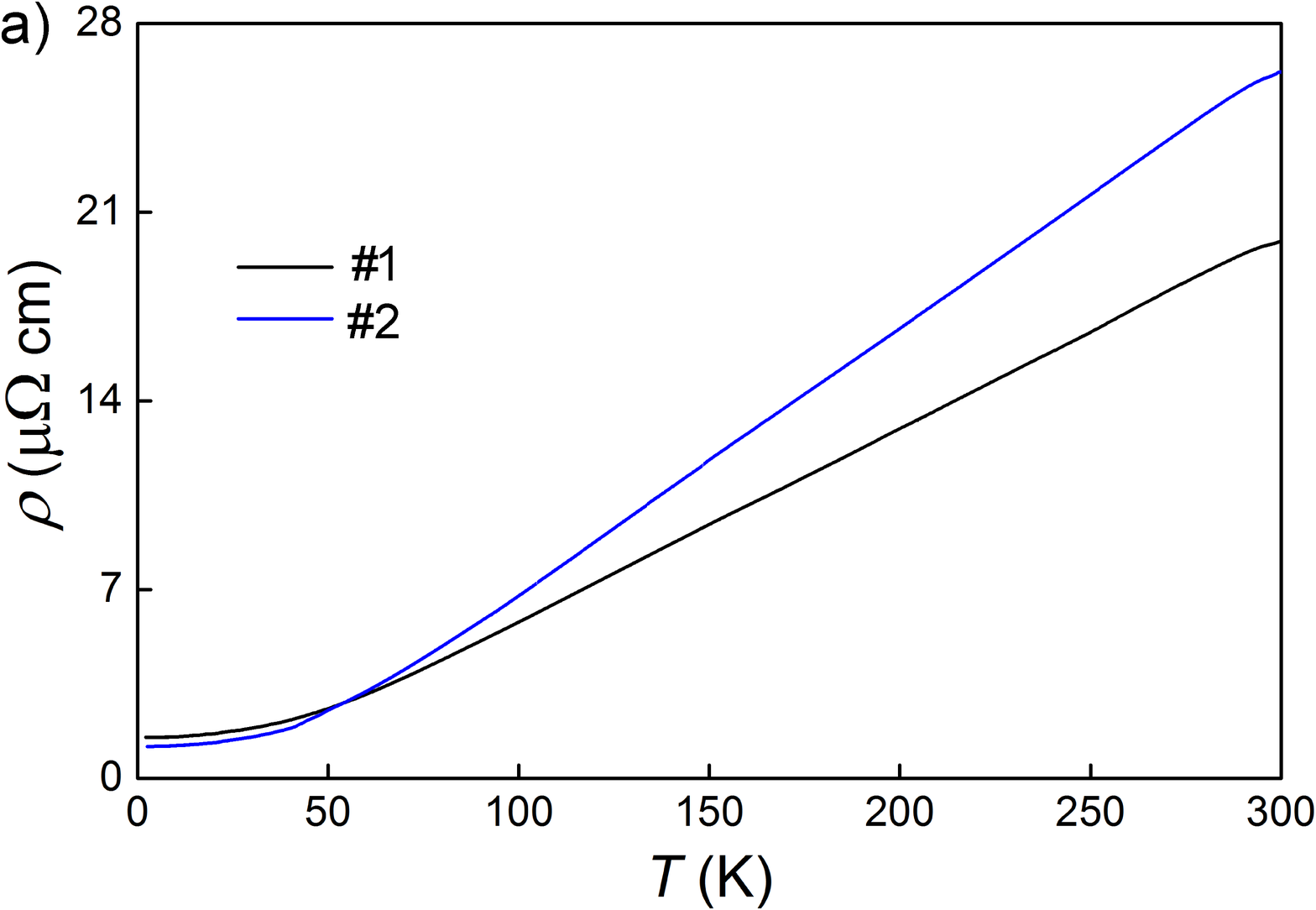}\\
\includegraphics[width=10cm]{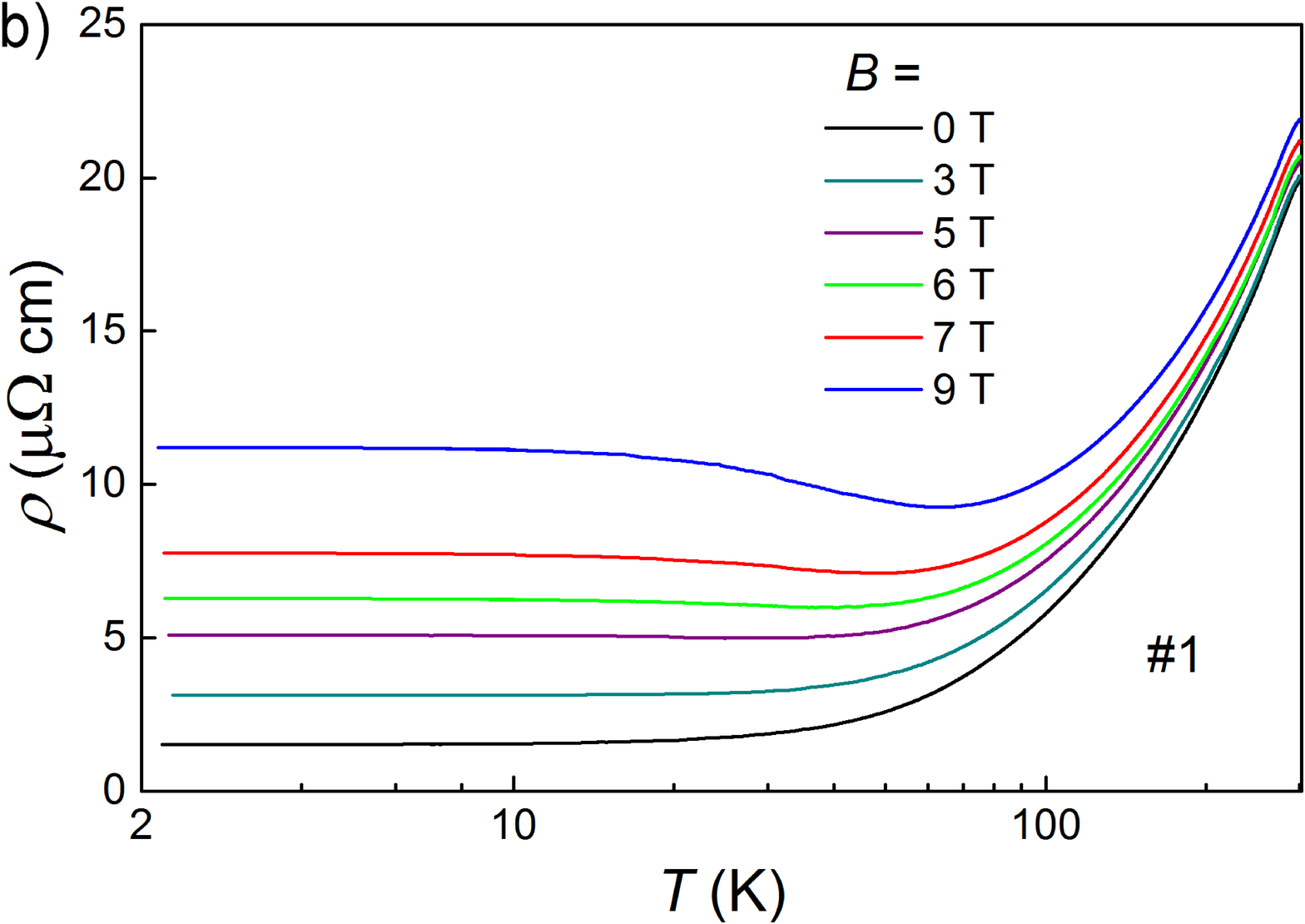}\\
\textbf{Figure S1.} (a) Resistivity of YSb versus temperature for samples \#1 and \#2.  (b) Resistivity of sample \#1 versus temperature, measured in different magnetic fields.   
\label{rho}
\end{figure}
\newpage
\rfoot{Supp. Mat. page 2 of 3}
\section*{Determination of effective masses}
Observation of clear SdH oscillations at different temperatures allowed us to determine effective masses of charge carriers responsible for their strongest components. Fitting of FTT amplitudes for components denoted as $\alpha$ and $\beta$ with $R_{i(=\alpha,\beta)}(T)=(\lambda m^*_iT/B)/\sinh(p\lambda m^*_iT/B)$  function with $B$ set at 9\,T and constant $\lambda=2\pi^2k_Bm_0/e\hbar\:(\approx14.7\;$T/K), as shown in Fig.\,S2, yielded effective masses: $m^*_\alpha=\,0.29m_0$ and $m^*_\beta=\,0.3m_0$.
\begin{figure}[h]
\includegraphics[width=10cm]{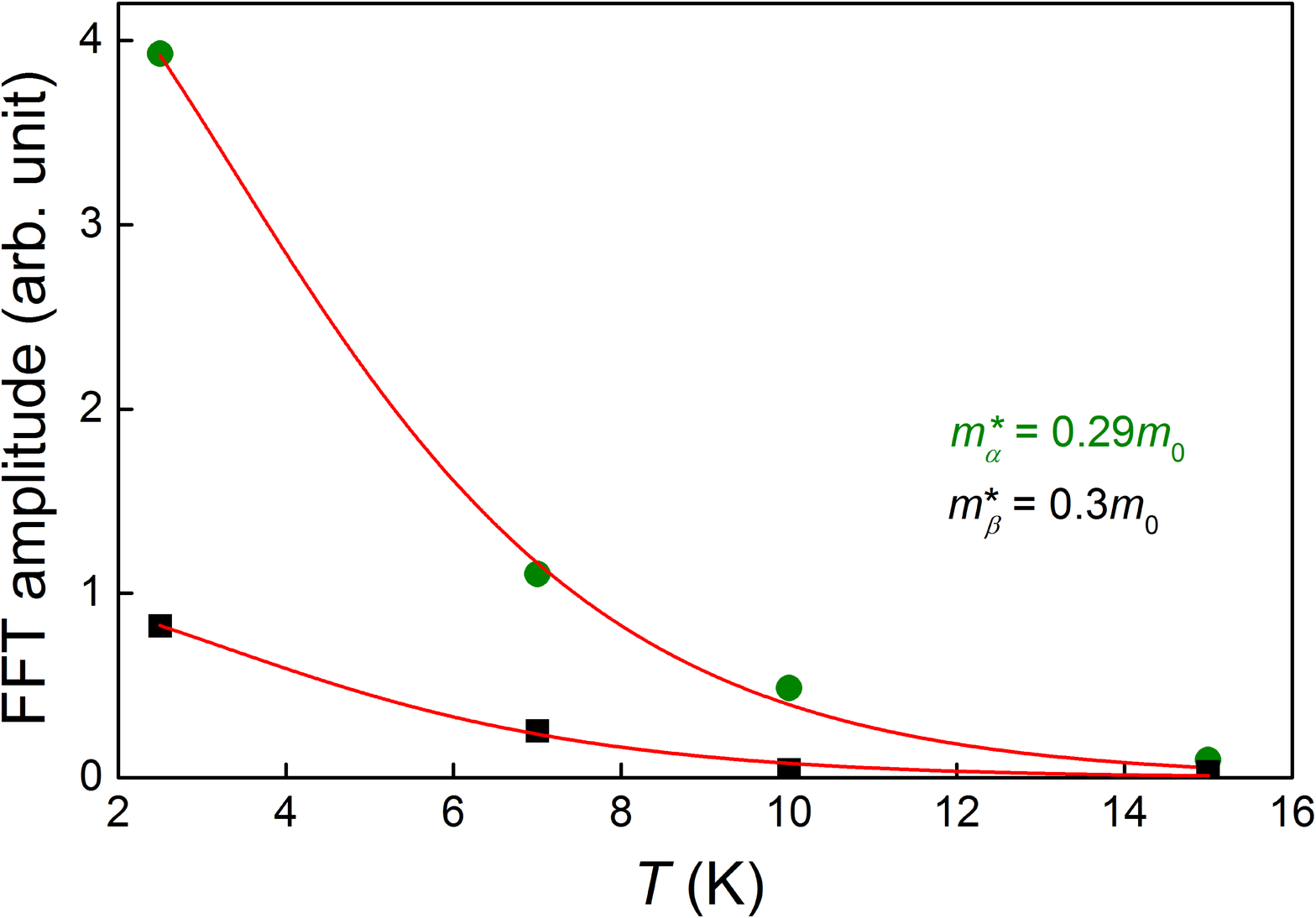}\\
\textbf{Figure S2.} Amplitudes of SdH oscillations corresponding to orbits $\alpha$ (circles) and $\beta$ (squares) obtained from FFT analysis. Red lines represent fits with function described in text yielding $m^*_\alpha=\,0.29m_0$ and $m^*_\beta=\,0.3m_0$. 
\label{rho}
\end{figure}
In their recent arXiv preprint Yu et al. reported $m^*_\alpha=\,0.17m_0$ and $m^*_\beta=\,0.27m_0$.\cite{Yu2016}

\section*{Multi-component Lifshitz-Kosevich function}
The oscillatory component of resistivity in magnetic field may be expressed as
\cite{Lifshitz1958,Shoenberg1984,Seiler1989}:\\\\
$$\Delta\rho_{xx} = \rfrac{5}{2}\sum_i\sqrt{\frac{B}{2p_if_i}} R_{T,i}R_{D,i}R_{S,i} \cos\bigg(2\pi\big(p_if_i/B-p_i\varphi_i-\frac{1}{8}\big)\bigg)=\qquad\qquad\qquad\qquad\qquad\qquad\qquad$$
$$\qquad\qquad\qquad= \rfrac{5}{2}\sum_i  \sqrt{\frac{p_i}{2f_iB}}
\frac{\lambda m^*_iT\:\exp(-p_i\lambda m^*_iT_{D,i}/B)\cos(p_i\pi m^*_ig^*_i)}
{\sinh(p_i\lambda m^*_iT/B)}\:
\cos\bigg(2\pi\big(p_if_i/B-p_i\varphi_i-\frac{1}{8}\big)\bigg),\qquad\rm{(S1)}$$\\
where for $i$-th SdH component: \\
\begin{tabular}{l l}$f_i\quad$ is frequency, \\
$R_{D,i}=\exp(-p_i\lambda m^*_iT_{D,i}/B)$ &-- the Dingle reduction factor, \\$R_{T,i}=(p_i\lambda m^*_iT/B)/\sinh(p_i\lambda m^*_iT/B)$& -- the temperature reduction factor and \\$R_{S,i}=\cos(p_i\pi m^*_ig^*_i)$ &-- the spin factor.\end{tabular}\\
$\varphi_i$ is the phase, $p_i$ denotes harmonic number , $m^*_i$ cyclotron mass (in units of free electron mass $m_0$), $g^*_i$ effective g-factor and $T_{D,i}$ the Dingle temperature.  
$\lambda=2\pi^2k_Bm_0/e\hbar\:(\approx14.7\;$T/K) is constant. 
\\\\
In order to fit complex $\Delta\rho_{xx}$ data shown in Fig.\,6 without preassuming the harmonic numbers of components we used a simplified version of the Equation S1:
$$\qquad\qquad\Delta\rho_{xx} = \sum_{i} a_i\sqrt{1/B}\:
\frac{\exp(-c_i/B)}{\sinh(b_i/B)}\:
\cos\bigg(2\pi\big(f_i/B-\varphi_i-\frac{1}{8}\big)\bigg).\qquad\qquad\qquad\qquad\qquad\qquad\qquad\qquad\rm{(S2)}$$
\newpage 
\rfoot{Supp. Mat. page 3 of 3}
Here $\exp(-c_i/B)$ represents $R_{D,i}$, whereas 
$a_i\sqrt{1/B}/\sinh(b_i/B)\;[\:\propto \sqrt{B}\:R_{T,i}R_{S,i}]$ comprises $R_{T,i}$ and $R_{S,i}$ (with $b_i=p_i\lambda m^*_iT$ and $c_i=p_i\lambda m^*_iT_{D,i}$). 
\\
Initial fit revealed that $\varphi_2=2\varphi_1$ and $\varphi_4\approx3\varphi_1$, thus we could identify second and fourth component as second and third harmonic, respectively, of the strongest oscillation with $\varphi_1=360\,$T. \\
Accordingly, we put constraints on parameters: $b_2=2b_1,\:b_3=3b_1,\:c_2=2c_1$ and $c_3=3c_1$. 
\\
\\  
For $T=2.5\,$K (temperature of our measurement), $m^*_1=0.29$ and $m^*_3=0.3$ (since $m^*_\alpha=0.29m_0$ and $m^*_\beta=0.3m_0$, as it was shown above) one obtains $b_1=11.51\,$T and  $b_3=11.9\,$T. 
\\When $b_1$ and $b_2$ were fixed at these values, the final fit yielded parameters collected in the Table below: 
\\\\{\bf Supplementary Table}~~Parameters obtained from fit of the multi-component Lifshitz-Kosevich function to data-points shown in Fig.\,6 of the main text. Effective masses and Dingle temperatures calculated from $b_i$ and $c_i$ parameters are also shown.
\\\\
\begin{tabular}{l *{6}{c}} \hline\hline\\ 
{~~~~$i=$}& 1 & 2 & 3 & 4 & 5 & 6 \\
\hline\\
$f_i\!\:\;$(T)& 360 & 720 & 740 & 1072 & 1160 & 1430\\\\
FS-sheet assignment: &$\alpha$&$2\alpha$&$\beta$&$3\alpha$&$\delta$&$\alpha_1$ \\\\
$\varphi_i$ & 0.74(3)& 0.50(4) & 0.74(2) & 0.60(2) & 0.68(3) & 0.31(2)\\\\
$a_i$ & 0.41 & 0.41 & 0.11 & 0.41 & 0.005 & 0.0037 \\\\ 
$b_i$ & 11.51 & 23.02 & 11.91 & 34.53 & 30 & 40 \\\\
$m^*_i\;\:$($m_0$) & 0.29 & -- & 0.3 & -- & 0.76 & 1.00 \\\\
$c_i$ & 4.56 & 9.11 & 22.93 & 13.67 & 54.24 & 64.00 \\\\
$T_{D,i}$ (K) & 1.07 & -- & 5.2 & -- & 4.9 & 4.35 \\
\hline\hline
\end{tabular}
\\\\
The ratio $m^*_6/m^*_1(= m^*_{\alpha_1}/m^*_{\alpha})= 3.48$ is the anisotropy of effective mass of electrons on the $\alpha$-sheet of FS. This value is in excellent agreement with values of 3.4 and 3.6 derived from angular behavior of $MR$ and SdH oscillations, respectively.   
\setcounter{page}{12}
\end{document}